# REVERSIBLE AND IRREVERSIBLE PROCESSES IN DISPERSIVE/DISSIPATIVE OPTICAL MEDIA
## Electro-magnetic free energy and heat production

C. Broadbent, G. Hovhannisyan, M. Clayton, J. Peatross, S. A. Glasgow[a]

## 1. INTRODUCTION

Experiments involving slow[1] and fast[2-7] light have motivated various theoretical investigations[8,9] that interpret the phenomena arising from dispersion and dissipation in optical systems. Previously, we showed that the interpretation of these phenomena often referred to in the literature were intrinsically associated with incomplete energy accounting.[9] Specifically, we showed that for a dissipative dielectric, 1) there is an unambiguous notion of the total energy of the medium-field system, and 2) this total energy can in nowise demonstrate "fast" dynamics, neither locally nor globally.

This otherwise satisfying result is tempered by the realization that the total energy contains a component which is "dynamically unrecoverable," and should be viewed either as a static component of a closed system, or as part of a continual loss from an open system. In a closed system the "dynamically unrecoverable" energy is like a trail of bricks left behind a wagon; stationary pieces that are still part of the wagon-brick system. In the open system we think of the dropped bricks as having returned to the earth; once dropped they are no longer considered part of the wagon-brick system. Either viewpoint cheapens the result that total energy transport is luminal; one might suppose that luminality is only a consequence of a component which is either not moving at all or is completely lost from the system's energy accounting and whose transport properties should not be considered in the first place. In this work we do not address the question of luminality of the dynamically recoverable energy. Instead, we consider the heretofore-unsolved problem of unambiguously separating the recoverable from the unrecoverable energy in a general linear dielectric medium.

---

[a] S. A. Glasgow, G. Hovhannisyan, and M. Clayton, Department of Mathematics, J. Peatross and C. Broadbent, Department of Physics, Brigham Young University, Provo, Utah, 84602.

We use ideas from Landau and Lifshitz [LL][10] and Oughstun and Sherman [OS][11] by which the static component of energy is interpreted as the heat dissipated to the medium. [LL] explicitly make this interpretation, in the case of an arbitrary medium and arbitrary process, only in the limit of large time, while [OS] indicate that such a notion should exist at any time. In this sense [OS] consider a notion of heat production that is manifestly dynamical. [LL] establish a dynamical notion of *internal energy* (the sum of heat and free energy; the total energy), but they quickly restrict to *the non-dispersive case*, and proceed to define it as the difference between (some fixed notion of) the medium-field energy with and without the history of the instigating fields. They deem this type of definition as being thermodynamically significant. Of course, as they point out, the non-dispersive case is also the (macroscopically) conservative case. Consequently in the case they consider there can be no notion of heat dissipated, neither dynamically nor asymptotically. Therefore the total energy is just the free energy. Moreover [LL] continue by indicating that such a thermodynamically significant notion of (dynamical) internal energy probably cannot exist for a dissipative system. From the open system viewpoint this postulate is valid. Returning to the brick-wagon analogy, if one stops accounting for lost bricks, then the total amount of bricks in the brick wagon system is constantly changing; bricks are no longer conserved. Consequently, with no notion of dynamical internal energy, they do not prescribe a meaningful notion of its components, free and heat energy.

In this paper we define the dynamical free and heat energy in a stationary, causal, and purely dissipative medium (a *simple* medium). We show how the definition 1) vindicates a certain component of the viewpoint of [OS], namely that there is a meaningful notion of dynamical heat production, 2) creates a thermodynamically significant notion of total energy, generalizing the [LL] notion of internal energy, and 3) makes connections with intuitive properties for the simple Lorentz model and validates the interpretation of those properties.

## 2. REVERSIBLE AND IRREVERSIBLE ENERGY

From Poynting's theorem, the well-known total energy density in the system is

$$u(t) := \frac{1}{2} \mathrm{E}^2(t) + \frac{1}{2} \mathrm{H}^2(t) + \int_{-\infty}^{t} \mathrm{E}(\tau) \dot{\mathrm{P}}(\tau) d\tau, \tag{1}$$

where we have restricted to the scalar case, in order to simplify the following discussion.

The first two terms in Eq. (1) give the electromagnetic field energy density, and the last term represents the energy density stored in the coupled medium-field system. We refer to this third component as the *interaction energy* and denote it as $u_{\mathrm{int.}}(t)$ (previously we called this the exchange energy[12]).

The change in the free energy in a reversible isothermal process is defined as the difference between the change in the internal energy and the change in the heat energy during this process. An immediate consequence of this definition is that this change in free energy is equivalent to the work done on the system. Since the process is reversible, this work done by some external agent on the system is equivalent to the work the system can do on, or return to, the agent. We generalize this idea to irreversible/dissipative/non-equilibrium processes by defining the *dynamical free energy* in all cases to be the work the system can do on, or return to, an external agent. This generalized notion of free energy thus defined reduces to the classical notion in the reversible case.

It can be shown that the first two terms in Eq. (1) always constitute energy (density) that the medium-field system can return to a causal agent. Thus these first two terms always constitute a component of the dynamical free energy just defined. Consequently our search for the total dynamical free energy can be restricted to a discussion of the third term, the interaction energy $u_{\text{int.}}(t)$. Though our generalization of the free energy is manifestly dynamical, in the following we will abbreviate and refer to this non-equilibrium object as, simply, the free energy.

[OS] claim that $u_{\text{int.}}(t)$ should in principle be decomposed into two pieces, one of which represents energy stored in the medium "reactively" (their term), the other of which represents energy stored "latently" (our term). The free energy is then the sum of the field energy term and the reactive energy term. The heat energy corresponds to the latent term of the interaction energy. We will define and then find the reactive and latent energy terms.

Utilizing a theorem from our previous work,[9] we first rewrite the interaction energy

$$u_{\text{int.}}(t) = \int_{-\infty}^{+\infty} \omega \text{Im}[\chi(\omega)] \left| \hat{\text{E}}_t(\omega) \right|^2 d\omega, \tag{2}$$

where the *instantaneous* (or *causal*) *spectrum* $\hat{\text{E}}_t(\omega)$ is defined by

$$\hat{\text{E}}_t(\omega) := \frac{1}{\sqrt{2\pi}} \int_{-\infty}^{t} \text{E}(\tau) e^{i\omega\tau} d\tau. \tag{3}$$

and $\chi(\omega)$ is the susceptibility of the medium: $\hat{\text{P}}(\omega) = \chi(\omega)\hat{\text{E}}(\omega)$.

Manifestly $u_{\text{int.}}(-\infty) = 0$. We note that by the work of [LL], $u_{\text{int.}}(+\infty)$, the [LL] asymptotic heat, is non-negative for all fields interacting with any simple medium. Equation (2) generalizes this result and shows that a passive dielectric also has the property that the energy of interaction never runs a deficit at any finite time: $u_{\text{int.}}(t) \geq 0$ for any finite $t$. This is to say that, at any given time $t$, the field has done more work on the medium than the medium has done against the field. This property is important in order to make an unambiguous, dynamical separation of this term into reactive and latent energy pieces.

A physically relevant and unambiguous way to separate the interaction energy at each time $t$, for any field $\text{E}(t)$ and any susceptibility (of a simple dielectric), into "reactive" and "latent" pieces, is to reinterpret "reactive," to mean *reversible*. We make this notion precise by extending a well-accepted interpretation. In the [LL] interpretation $u_{\text{int.}}[\text{E}](+\infty) = u_{\text{int.}}[\text{E}](+\infty) - 0 = u_{\text{int.}}[\text{E}](+\infty) - u_{\text{int.}}[\text{E}](-\infty)$ represents the energy dissipated or *lost* to the medium via the signal $\{\text{E}(\tau) | -\infty < \tau < +\infty\}$ since time $t = -\infty$. Consequently, $u_{\text{int.}}[\text{E}](+\infty) - u_{\text{int.}}[\text{E}](t)$ represents the energy lost to the medium that is subject to the future signal $\{\text{E}(\tau) | t < \tau < +\infty\}$ since time $t$, given a preparation of the medium established by the past signal $\{\text{E}(\tau) | -\infty < \tau < t\}$ prior to time $t$. When this difference of energies is negative, we should refer to its opposite, $u_{\text{int.}}[\text{E}](t) - u_{\text{int.}}[\text{E}](+\infty)$, which is positive, as energy that, due to the future field, is returned to the field and so comprises a *reversible process* of borrowing energy from the field and returning it at a later time ($t = \infty$). Obviously, given any fixed past field interacting with a medium, some future fields yield larger energies in the borrow-return reversible process than others. Mathematically, the difference $u_{\text{int.}}[\text{E}](t) - u_{\text{int.}}[\text{E}](+\infty)$ is larger for some future fields than others. This idea suggests a definition.

Definition. The *reversible energy* $u_{rev.}[E](t)$ *of the signal* $\{\text{E}(\tau) | -\infty < \tau < t\}$ *at time t in the medium* is the supremum[b] of values $u_{\text{int.}}[\text{E}](t) - u_{\text{int.}}[\text{E}](+\infty)$ can obtain over alternative futures, i.e. over alternative future signals $\{\text{E}(\tau) | t < \tau < +\infty\}$.

We can rewrite this conceptual definition operationally as,

$$u_{rev.}[\text{E}](t) := \sup_{\text{E}_f} \left\{ u_{\text{int.}}[\text{E}H_t^- + \text{E}_f](t) - u_{\text{int.}}[\text{E}H_t^- + \text{E}_f](+\infty) \right\}, \qquad (4)$$

where

$$H_t^-(\tau) := \begin{cases} 1, & \tau < t; \\ 0, & \text{otherwise}, \end{cases} \qquad \text{E}_f(\tau) = 0, \ \tau < t, \qquad (5)$$

and $\text{E}_f$ denotes the possible future fields, $\{\text{E}(\tau) | t < \tau < +\infty\}$. Since the interaction energy is causal, i.e. since at any given time $t$ it depends only on values of the field prior to $t$, the first term in Eq. (4) does not depend on the future field $\text{E}_f$. Therefore, we can simplify Eq. (4) to

$$u_{rev.}[\text{E}](t) := u_{\text{int.}}[\text{E}](t) - \inf_{\text{E}_f} u_{\text{int.}}[\text{E}H_t^- + \text{E}_f](+\infty). \qquad (6)$$

Thus, from the computational point of view, the complement of this component of the total interaction energy is more fundamental. For the time being we will call this complement the irreversible energy:

---

[b] The supremum (sup) and infemum (inf) are the least upper bound and greatest lower bound, respectively. For example, the sequence (1, .5, .25, .125, .0625, … ) has infemum equal to zero though no minimum exists. The supremum and infemum, however, are the maximum and minimum, respectively, of a series if the maximum or minimum exist.

$$u_{irrev.}[\mathrm{E}](t) := \inf_{\mathrm{E}_\mathrm{f}} u_{int.}[\mathrm{E}H_t^- + \mathrm{E}_\mathrm{f}](+\infty). \tag{7}$$

Defined this way, $u_{irrev.}[\mathrm{E}](t)$ lends itself easily to the calculus of variations. Noting that any solution $\mathrm{E}_\mathrm{f}(\tau)$ vanishes for all times $\tau < t$ (Eq. 5), the Fourier transform of the concatenated field $(\mathrm{E}H_t^- + \mathrm{E}_\mathrm{f})(\tau)$

$$\mathsf{F}\,[\mathrm{E}H_t^- + \mathrm{E}_\mathrm{f}](\omega) = \frac{1}{\sqrt{2\pi}} \int_{-\infty}^{t} \mathrm{E}(\tau) e^{i\omega\tau} d\tau + \frac{1}{\sqrt{2\pi}} \int_{t}^{\infty} \mathrm{E}_\mathrm{f}(\tau) e^{i\omega\tau} d\tau \tag{8}$$

is given by $\hat{\mathrm{E}}_t(\omega) + \hat{\mathrm{E}}_\mathrm{f}(\omega)$. Then, by using Eq. (2), and setting the variational derivative over $\mathrm{E}_\mathrm{f}(\tau)$ to be zero, we have

$$\delta_{\mathrm{E}_\mathrm{f}} u_{int.}[\mathrm{E}H_t^- + \mathrm{E}_\mathrm{f}](\infty) = \delta_{\hat{\mathrm{E}}_\mathrm{f}} \int_{-\infty}^{+\infty} \omega \mathrm{Im}[\chi(\omega)] \left|\hat{\mathrm{E}}_t(\omega) + \hat{\mathrm{E}}_\mathrm{f}(\omega)\right|^2 d\omega = 0. \tag{9}$$

Introducing new functions $\mathrm{E}_+(\tau)$ and $\mathrm{E}_-(\tau)$ by shifting the fields in the time domain by $t$, i.e. $\mathrm{E}_\mathrm{f}(\tau) = \mathrm{E}_+(\tau - t)$, and $(\mathrm{E}H_t^-)(\tau) = \mathrm{E}_-(\tau - t)$, shifts the transforms according to $\hat{\mathrm{E}}_\mathrm{f}(\omega) = e^{i\omega t}\hat{\mathrm{E}}_+(\omega)$ and $\hat{\mathrm{E}}_t(\omega) = e^{i\omega t}\hat{\mathrm{E}}_-(\omega)$. Since the extra phase factors $e^{i\omega t}$ can be neglected because of the modulus in Eq. (9), we can replace $\hat{\mathrm{E}}_\mathrm{f}(\omega)$ and $\hat{\mathrm{E}}_t(\omega)$ with $\hat{\mathrm{E}}_+(\omega)$ and $\hat{\mathrm{E}}_-(\omega)$, which, importantly, are functions analytic in the upper and lower half planes, respectively. Finally, we take the variation over the new quantity $\hat{\mathrm{E}}_+(\omega)$, and using the symmetries of the Fourier transform of a real-valued time-series, we get

$$2 \int_{-\infty}^{+\infty} \omega \mathrm{Im}[\chi(\omega)] \left(\hat{\mathrm{E}}_-(\omega) + \hat{\mathrm{E}}_+(\omega)\right) \delta\hat{\mathrm{E}}_+^{\,*}(\omega) d\omega = 0. \tag{10}$$

Since $\hat{\mathrm{E}}_+(\omega)$ is analytic in the upper half plane, so is its variation $\delta\hat{\mathrm{E}}_+(\omega)$. Consequently, $\delta\hat{\mathrm{E}}_+^{\,*}(\omega)$ is analytic in the lower half plane. A certain subset of these variations are *admissible*. However, we will briefly postpone discussion of this requirement.

Because $\delta\hat{\mathrm{E}}_+^{\,*}(\omega)$ is analytic in the lower half plane, but is otherwise arbitrary, Eq. (10) is impossible unless the rest of the integrand is also analytic in the lower half plane. If the entire integrand is analytic and rapidly vanishing in the lower half plane, we can obtain the integral in Eq. (10) by extending the contour with an arbitrarily large half-circle in the lower half plane. From Cauchy's theorem, the integral over the whole contour will be zero. Because the integrand is rapidly vanishing at infinity, the integral over the half-circle will go to zero as we extend the radius to infinity. This ensures that, regardless of the form of an admissible variation, that the integral over the real axis (in Eq. (10)) is zero. By defining a new quantity $\hat{Z}_-(\omega)$, we can rewrite this condition on $\hat{\mathrm{E}}_+(\omega)$ as

$$\hat{Z}_-(\omega) = \omega \mathrm{Im}[\chi](\omega) \left(\hat{\mathrm{E}}_-(\omega) + \hat{\mathrm{E}}_+(\omega)\right), \tag{11}$$

where $\hat{Z}_-(\omega)$ is required to be analytic and rapidly vanishing in the lower-half complex plane.

Equation 11 is the main result of this paper. Its solution allows for the separation of the interaction energy into reversible and irreversible components for *any* simple medium. Notably, Eq. (11) constitutes one equation with two unknowns, $\hat{Z}_-(\omega)$ and $\hat{E}_+(\omega)$, and is only solvable by utilizing important constraints inherent in the original problem. First, passive susceptibilities are analytic in the upper-half complex plane, have the asymptotics such that for large real frequencies $\omega$,

$$\text{Re}[\chi(\omega)] = \mathrm{O}(\frac{1}{\omega^2}),\ \text{Im}[\chi(\omega)] = \mathrm{O}(\frac{1}{\omega^3}),\ \text{Im}[\chi(\omega)] \neq \mathrm{o}(\frac{1}{\omega^3}),\ \omega \to \infty, \tag{12}$$

and for small real frequencies $\omega$,

$$\text{Im}[\chi(\omega)] = \mathrm{O}(\omega),\ \text{Im}[\chi(\omega)] \neq \mathrm{o}(\omega),\ \omega \to 0, \tag{13}$$

and have the property that $\omega\text{Im}[\chi(\omega)] \geq 0$ for all real frequencies, with equality only when $\omega = 0$. These asymptotics put restrictions on admissible solutions to this problem via the requirement that the associated interaction energy be finite. Second, $\hat{E}_+(\omega)$ is analytic in the upper-half complex plane, and $\hat{E}_-(\omega)$ and $\hat{Z}_-(\omega)$ are analytic in the lower-half complex plane. The general solution to Eq. (11), together with the constraints just mentioned, is found by recasting the problem in terms of a classical one from complex analysis: the Riemann-Hilbert problem.[13] The solution is (the Fourier transform of) an optimal future field (which we will call the reversal field and denote as $\{E_{rev.}[E,t](\tau)|\ t \leq \tau < \infty\}$) that will give the infemum [LL] asymptotic heat associated with some past field $\{E(\tau)|-\infty < \tau < t\}$; it is a future field that uniquely specifies $u_{irrev.}[E](t)$, and, by it's complimentary nature, $u_{rev.}[E](t)$. In section four we show in general that the irreversible energy is consistent with thermodynamic notions of heat; the irreversible energy is shown to never decrease. We also show that the reversible energy (the difference between the interaction energy and the irreversible energy) is consistent with thermodynamic notions of free energy.

## 3. SOLUTION FOR A SINGLE LORENTZ OSCILLATOR

In this section we utilize the definition for reversible and irreversible energy to find the reversal field, and consequently, the reversible and irreversible energies for a single species Lorentz oscillator. While the quantities associated with free energy and heat have been known for a dissipative Lorentz oscillator model of a medium since, at least, the work of Loudon in 1970,[14] we pursue the trivial Lorentz model to reinforce Loudon's interpretation for a well-known model of a medium. However, we have found the intuitive method used by Loudon for the single Lorentz oscillator model *cannot* determine the free and heat energy in general. That method involves determining the kinetic and po-tential energies of the components of the system and calling the sum of those free energy, and ascribing all other forms to energy dissipation. We have found, for the next simplest case of a double species Lorentz oscillator, for all times $t$ (except infinity) the kinetic and potential energies are always greater than the reversible energy. As will be examined closely in an upcoming publication,[15] in which we consider general (non-Lorentzian) media, the reason for this is the effective irreversibility of (microscopic) decoherence among system elements.

Consider the single Lorentz oscillator with susceptibility $\chi(\omega)$,[c]

$$\chi(\omega) = \frac{-\omega_p^2}{(\omega - (\omega_0 - i\gamma))(\omega - (-\omega_0 - i\gamma))}. \tag{14}$$

In contrast to the general method of solution, for a single Lorentz oscillator it is easier to use a Wiener-Hopf factorization method than the theory of the Riemann-Hilbert problem to find the reversal field (the solution to Eq. (11) with associated constraints). Once factored, the theorem from complex analysis dictating that bounded entire functions are constants gives the solution. The entire derivation will not be presented in this paper. Instead of solving for $\hat{E}_+[E,t](\omega)$ in Eq. (11) it is necessary to solve for $\omega\hat{E}_+[E,t](\omega)$. After some work, we find that

$$\omega\hat{E}_+[E,t](\omega) = -\left(\text{Re}\left[\hat{E}_-(\omega_0 - i\gamma;t)\right] + \frac{\gamma}{\omega_0}\text{Im}\left[\hat{E}_-(\omega_0 - i\gamma;t)\right]\right)\omega - \frac{i(\omega_0^2 + \gamma^2)}{\omega_0}\text{Im}\left[\hat{E}_-(\omega_0 - i\gamma;t)\right], \tag{15}$$

where

$$\hat{E}_-(\omega_0 - i\gamma;t) := \frac{1}{\sqrt{2\pi}}\int_{-\infty}^{0} E(t+\tau)e^{i(\omega_0 - i\gamma)\tau}d\tau. \tag{16}$$

Then

$$\omega\hat{E}_+[E,t](\omega) = C_1\omega + C_0, \tag{17}$$

where, to simplify the following discussion, $C_n$ is the coefficient of $\omega^n$ in Eq. (15).

Given that $\omega\hat{E}_+[E,t](\omega)$ does not generically have a zero at $\omega = 0$ (i.e. $C_0 \neq 0$), Eq. (17) does not define $\hat{E}_+[E,t](\omega)$ as an analytic function in the upper-half $\omega$ plane.

---

[c] We note that the denominator of the susceptibility given in Eq. (14) differs from the more common use of $(\omega_0^2 - i\gamma\omega - \omega^2)$ which emphasizes resonance and damping terms. The denominator in Eq. (14) proves more useful to this work in that it explicitly denotes the placement of the poles of the susceptibility in the complex plane.

Consequently the latter's inverse transform, the function $\tau \mapsto \mathrm{E}_+[\mathrm{E},t](\tau)$, will not vanish for times $\tau < 0$, and finally the reversal field $\tau \mapsto \mathrm{E}_{rev.}[\mathrm{E},t](\tau)$, will not vanish for times $\tau < t$. This problem is generated by the fact that the [LL] asymptotic interaction energy (from which the irreversible energy is computed using the past/reversal concatenated field as in $u_{int.}[\mathrm{E}H_t^- + \mathrm{E}_{rev.}](+\infty)$) relies solely on the derivatives of the incident field; it can be rewritten like this,

$$u_{\mathrm{int.}}(+\infty) = \int_{-\infty}^{+\infty} \frac{\mathrm{Im}[\chi(\omega)]}{\omega} \left| -i\omega \hat{\mathrm{E}}(\omega) \right|^2 d\omega \tag{18}$$

and thus, cannot distinguish between fields that vary by a constant. However, this problem is easily remedied. Instead of solving for $\hat{\mathrm{E}}_+[\mathrm{E},t](\omega)$ in Eq. (17), we map $\omega \to \omega + i\varepsilon$, and then solve for $\hat{\mathrm{E}}_+[\mathrm{E},t](\omega + i\varepsilon)$

$$\hat{\mathrm{E}}_+[\mathrm{E},t](\omega + i\varepsilon) = C_1 + \frac{C_0}{\omega + i\varepsilon}, \tag{19}$$

where $\varepsilon$ denotes a positive infinitesimal. In this way $\hat{\mathrm{E}}_+[\mathrm{E},t](\omega + i\varepsilon)$ becomes distributional and, most importantly, becomes analytic in the upper-half plane. Since the distribution $1/(\omega + i\varepsilon)$ can be decomposed into two distributions, namely a delta function and a principal value distribution, $\hat{\mathrm{E}}_+[\mathrm{E},t](\omega + i\varepsilon)$ can be organized into three types of terms: a constant, a delta function, and a principal value distribution. In time $\tau$ these three become, respectively, a delta function (supported at $\tau = 0$), a constant (i.e. DC field), and a "sign" function, i.e. a piece-wise constant function returning the sign of its argument times 1. The amplitudes of the latter two components, both arising from the distribution $1/(\omega + i\varepsilon)$, are such that $\mathrm{E}_+[\mathrm{E},t;\varepsilon](\tau)$ vanishes for $\tau < 0$, and so

$$\mathrm{E}_{rev.}[\mathrm{E},t;\varepsilon](\tau) = \sqrt{2\pi}\left(C_1\delta(\tau-t) - iC_0 e^{-\varepsilon(\tau-t)}H_\tau^-(t)\right). \tag{20}$$

Equation (20) defines a family, in $\varepsilon$, of reversal fields for each time $t$. With the reversal field for each positive $\varepsilon$, there is an associated [LL] asymptotic heat. The infemum of these [LL] asymptotic heats, as $\varepsilon$ goes to zero, is the irreversible energy at time $t$. As $\varepsilon$ approaches arbitrarily close to zero the [LL] asymptotic heat (for the past/quasi-reversal concatenated field) approaches arbitrarily close to the infemum that defines the irreversible energy. Figures 1(b) and 1(e) on the following page emphasize that reversal fields with smaller $\varepsilon$ yield [LL] asymptotic heats that approach closer to the irreversible energy. However, as $\varepsilon$ gets smaller, it takes longer to realize the [LL] asymptotic heat for the past/quasi-reversal concatenated field. Figures 1(c) and 1(f) show that the interaction energy for a past/quasi-reversal field with smaller $\varepsilon$ takes longer to approach its final asymptotic value.

Consequently, for $\varepsilon$ arbitrarily close to zero, time must run arbitrarily close to infinity to realize the evolved heat. This means that when actually finding the irreversible energy, care must be taken as to when the limit of $\varepsilon$ going to zero is evaluated. However, when calculating the irreversible energy (and by compliment, the reversible energy) according to

$$u_{irrev.}[\mathrm{E}](t) := u_{int.}\left[\mathrm{E}H_t^- + \mathrm{E}_{rev.}[\mathrm{E},t;\varepsilon]\right](+\infty), \tag{21}$$

by means of the interaction energy representation in Eq. (2), $\varepsilon$ can be sent to zero with impunity because time, in the interaction energy, has already run to infinity. This results in the following reversible and irreversible energies for the single Lorentz oscillator:

$$u_{rev.}[\mathrm{E}](t) = \pi \frac{\omega_p^2}{\omega_0^2} \mathrm{Re}[(\omega_0 - i\gamma)\hat{\mathrm{E}}_-(\omega_0 - i\gamma;t)]^2 + $$
$$\pi \frac{\omega_p^2}{\omega_0^2}(\omega_0^2 + \gamma^2)\mathrm{Im}[\hat{\mathrm{E}}_-(\omega_0 - i\gamma;t)]^2, \tag{22}$$

$$u_{irrev.}[\mathrm{E}](t) = \frac{4\pi\gamma\omega_p^2}{\omega_0^2} \int_{-\infty}^{t} \mathrm{Re}[(\omega_0 - i\gamma)\hat{\mathrm{E}}_-(\omega_0 - i\gamma;\tau)]^2 d\tau. \tag{23}$$

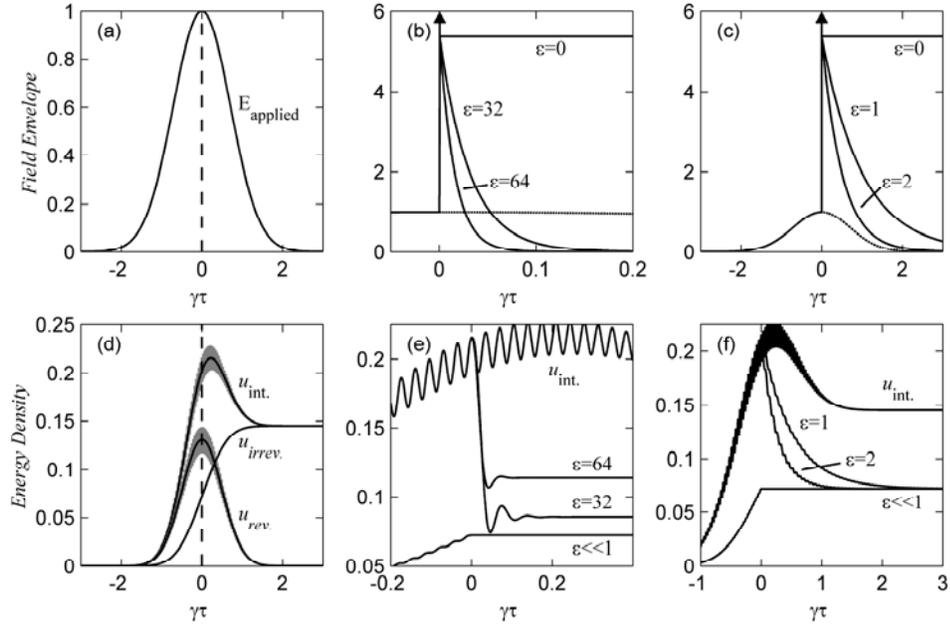

**Figure 1**. (a-c) Electric field envelopes in units of $E_0$. (a) An applied E-field. The dashed line in (a) and (d) denotes the time $\gamma t = 0.0005$. (b) Alternative future fields applied at $\gamma t = 0.0005$ taking the place of the applied E-field (dotted). Here $\varepsilon = 0$, 32 and 64. (c) Same as in (b) except $\varepsilon = 0$, 1, and 2. (d-f) energy in units of $E_0^2/2$. (d) Interaction, reversible, and irreversible energy for the applied E-field. Here grey depicts the actual rapid oscillations; the black is time-averaged. (e) Interaction energy for alternative future fields. Here $\varepsilon = 32$, 64. The top line is the interaction energy for the applied E-field. The bottom line gives the irreversible energy as it would have evolved for $\varepsilon \ll 1$. (f) Same as (e) except $\varepsilon = 1, 2$.

Manifestly, $u_{rev.}[\mathrm{E}](t)$, $u_{irrev.}[\mathrm{E}](t)$, and $\dot{u}_{irrev.}[\mathrm{E}](t) \geq 0$.

After a bit of work (involving the constitutive relation $\hat{\mathrm{P}}(\omega) = \chi(\omega)\hat{\mathrm{E}}(\omega)$ and Cauchy's theorem), it can be shown that

$$\mathrm{Im}[\hat{\mathrm{E}}_-(\omega_0 - i\gamma;t)] = -\frac{\omega_0}{\sqrt{2\pi}\omega_p^2}\mathrm{P}(t), \tag{24}$$

$$\mathrm{Re}[(\omega_0 - i\gamma)\hat{\mathrm{E}}_-(\omega_0 - i\gamma;t)] = \frac{\omega_0}{\sqrt{2\pi}\omega_p^2}\dot{\mathrm{P}}(t). \tag{25}$$

Substituting Eq. (24) and Eq. (25) back into Eq. (22) and Eq. (23), we find, verifying the interpretation of Loudon, that

$$u_{rev.}[\mathrm{E}](t) = \frac{1}{2\omega_p^2}\dot{\mathrm{P}}^2(t) + \frac{(\omega_0^2 + \gamma^2)}{2\omega_p^2}\mathrm{P}^2(t), \tag{26}$$

$$u_{irrev.}[\mathrm{E}](t) = \frac{2\gamma}{\omega_p^2} \int_{-\infty}^{t} \dot{\mathrm{P}}^2(\tau) d\tau. \tag{27}$$

Equation (26) shows that the reversible energy for a single Lorentz oscillator is the sum of its kinetic and potential energy. However, in general (i.e. for an arbitrary susceptibility), the reversible energy is not equal to the sum of the kinetic and potential energies of (microscopic) system elements, but is found, generically, to be less than that sum. As mentioned previously, this is due to the effective irreversibility of decoherence among system elements.

## 4. FREE ENERGY AND HEAT

We will now demonstrate that the definitions we have made for reversible and irreversible (interaction) energy not only have the properties their names suggest, but also always have the properties defining, classically, free energy and heat. Of course we have defined the first of these quantities in such a way as to establish it as the supremum on the amount of work the medium can do against the field and, so, since one defines (the change in) the free energy of a body thermodynamically as the work done on a body in a reversible isothermal process, the use of the term is justified from this semantic point of view. Better, however, is the fact that one can show that from the classic definition of free energy, in an irreversible isothermal process, volume also being held constant, free energy always decreases. The restrictions just stated amount to those that prohibit an introduction of total energy into the system. In the following, we will show that, after the instigating electric field has ceased subsidizing the reversible energy's existence, i.e. after the electric field ceases to introduce energy into the medium-field system, this component of the total energy can never increase and, moreover, decreases almost always.

The way that we will establish this result, however, is by showing first the easier fact that the irreversible energy never decreases and increases almost always *regardless of the state of the field* $E(t)$. In this way we establish this component of the energy as recording the irreversibility of the process defining the medium-field interaction, at least from an energetic point of view. This will be our justification for calling this component heat.

The proof that the irreversible energy cannot decrease requires no calculation, and follows directly from the definition of the irreversible energy, namely that, at a specific time $t$, it is the infemum of the interaction energy after this time: obviously if these dynamically calculated infima were to ever decrease after time $t$, the original (now candidate) value for the infemum at time $t$ would not have been an extrema after all. The only calculation-like statement that needs be made is that such an infemum is guaranteed to exist. This is assured by Eq. (2) which shows that the total interaction energy is positive definite, i.e. it can never obtain a value smaller than zero—see the paragraph after Eq. (2) alluding to this necessity. Having proved this fact regarding the dynamics of the irreversible energy, from now on we will call it the (dynamical) heat, and will denote it as $Q[E](t)$:

$$Q[E](t) := u_{irrev.}[E](t). \tag{28}$$

It is now clear that the reversible energy can never increase *after the field* $E(t)$ *has ceased*. Suppose that this field ceases at (i.e. is supported before) time $t_0$. Then, according to Eq. (6), (7), and now (28) for each time $t$ greater than $t_0$,

$$u_{rev.}[E](t) = u_{int.}[E](t_0) - Q[E](t), \tag{29}$$

an important feature of which is that the first term on the right is now a constant. From Eq. (29), it is now clear that the reversible energy decreases whenever the heat increases after time $t_0$. Having proved this fact regarding the dynamics of the reversible energy, from now on we will call it the mechanical energy (i.e. the component of the total energy associated with the medium field interaction having the characteristics of free energy), and will denote it as $M[\mathrm{E}](t)$:

$$M[\mathrm{E}](t) := u_{rev.}[\mathrm{E}](t) = u_{int.}[\mathrm{E}](t) - Q[\mathrm{E}](t). \tag{30}$$

## 5. SUMMARY

In this paper we have provided the essential definitions and concepts that allow for a general solution to the dynamical separation of free and heat energy first alluded to by [LL] in 1958, and later posed concretely by [OS] in 1994. We have vindicated a certain component of the viewpoint of [OS]: there is a meaningful notion of dynamical heat production. We have shown, contrary to the viewpoint of [LL], that there is a thermodynamically significant notion of internal (or total) energy for dissipative systems, and we have extended [LL]'s well known definition regarding evolved, asymptotic heat to pose a definition for the dynamical free energy and heat. This definition states that the amount of free energy in a system is the field energy plus the mechanical energy. In turn, the mechanical energy is defined as the supremum of all possible changes in the internal energy of the medium. Consequently, heat is equivalent to the infemum of all possible future internal energies. We have applied this definition to the single Lorentz oscillator model of a dielectric and have shown that it verifies the interpretation of Loudon involving an intuitive decomposition of dynamical free energy and heat. Lastly, we have shown for an arbitrary simple dielectric, that the definitions posed for dynamical free energy and heat yield the properties that characterize them classically.

In forth-coming works we will present the development of dynamical mechanical energy and heat for arbitrary simple dielectrics,[13] demonstrate that mechanical energy is almost always less than the sum of the kinetic and potential energies of the microscopic elements comprising the system,[15] and discuss the application of these concepts to super/subluminal phenomenon.[16]